\pdfoutput=1
\documentclass{eptcs}
\providecommand{\event}{ Workshop on Quantitative Formal Methods: Theory and Applications} 
\providecommand{\volume}{??}
\providecommand{\anno}{2009}
\providecommand{\firstpage}{1}
\providecommand{\eid}{??}
\usepackage{graphicx,tabularx,color,amsmath,verbatim,subfig, wasysym}

\title{Adaptive Scheduling of Data Paths using Uppaal Tiga%
\thanks{The research of Igna and Vaandrager was carried out in the context of the Octopus project, with partial support of the Netherlands Ministry of Economic Affairs under the Senter TS program.  This research was also supported by European Community's Seventh Framework Programme under grant agreement no 214755 (QUASIMODO).}}%
\author{Israa AlAttili \and Fred Houben \and Georgeta Igna \and Steffen Michels \and Feng Zhu \and Frits Vaandrager
\institute{Institute for Computing and Information Sciences\\
Radboud University Nijmegen\\
P.O.\ Box 9010, 6500 GL Nijmegen, The Netherlands}
}
\def\authorrunning{AlAttili, Houben, Igna, Michels, Zhu and Vaandrager}
\def\titlerunning{Adaptive Scheduling of Data Paths using Uppaal Tiga}

\begin{document}
\maketitle

\begin{abstract}
We apply Uppaal Tiga to automatically compute adaptive scheduling strategies for an industrial
case study dealing with a state-of-the-art image processing pipeline of a printer.
As far as we know, this is the first application of timed automata technology to an industrial
scheduling problem with uncertainty in job arrivals.
\end{abstract}

\section{Introduction}

Scheduling concerns the allocation of resources to activities over time in order to achieve some goals.
Scheduling problems occur in many different domains and a vast amount of research has been carried out in this area.
However, in the research literature scheduling is usually seen as a function of known, perfect inputs: the set of jobs,
their arrival times, the capacities of machines, the duration of activities, and other characteristics of the
problem are assumed to be known and static \cite{DB00}.
It has been observed that scheduling processes in practice are driven by uncertainty \cite{MBS89,MW99}.
This uncertainty may arise due to various sources (machine breakdown, unexpected arrival of new jobs, modification of existing jobs, uncertainty
of task durations,..).
McKay et al. \cite{MSB98} even claim that the dynamic characteristics of real-world scheduling environments render the bulk of
existing solution approaches for the job shop problem unusable when applied to practical problems.
The number of scientific publications devoted to scheduling in a setting with uncertainty is relatively small and most of them have
the flavor of AI planning rather than Operations Research \cite{DB00,AAM06}.
Altogether the problem of computing \emph{optimal} scheduling strategies in practical settings with uncertainty is largely open.
The present paper aims to address this problem.

Within the European AMETIST project \cite{AMETISTfinal}, steps have been taken in the development of a general theory
of scheduling inspired by the methodology of model checking and based on the timed automaton model of Alur and Dill \cite{AD94}.
In the AMETIST approach, components of a system are modeled as \emph{dynamical systems} with a state space and a well-defined dynamics.
All that can happen is expressed in terms of \emph{behaviors} that can be generated by the dynamical systems; these constitute the
semantics of the problem.
Verification, optimization, synthesis and other design activities explore and modify system structure so that the resulting behaviors are correct,
optimal, etc.
Using this approach, the project was able to derive schedules that were of comparable quality as those that were provided by an industrial
tool \cite{BehrmannBHM05}.
Most of the work within AMETIST did not address scheduling under uncertainty, with the notable exception of \cite{AAM06}, which studies
uncertainty in task durations.
Recently, however, an extension of the timed automata model checker Uppaal \cite{BehrmannDL04} has been proposed, called Uppaal Tiga, that
implements the first efficient on-the-fly algorithm for solving games based on timed game automata with respect to reachability
and safety properties \cite{tiga_manual,CDFLL05}.
Although still a prototype, we believe that this extension has potential as a tool for solving scheduling problems that involve uncertainty
(using the AMETIST approach).
In Uppaal Tiga, systems are specified through a network of timed game automata \cite{MPS95}.  These are timed automata in the sense of \cite{AD94}
where edges are marked as either controllable or uncontrollable.  This defines a two player game with on the one side the \emph{controller}
(mastering the controllable edges) and on the other side the \emph{environment} (mastering the uncontrollable edges).
Winning conditions of the game are specified through TCTL formulas and for instance state that, irrespective of the strategy used by
the environment player, the system player can always reach (or always avoid) certain states.
In a scheduling context, uncertainty can be modeled using uncontrollable edges.  Uppaal Tiga is then able to synthesize strategies
for controlling the system such that scheduling objectives are met irrespective of the timing of uncontrollable edges.

In order to demonstrate the practical usefulness of Uppaal Tiga for solving scheduling problems with uncertainty, we have applied the tool to an industrial
case study from Oc\'{e} Technologies that concerns the scheduling of a state-of-the-art image processing pipeline of a printer.
In \cite{Octopus08}, an initial version of this scheduling problem has been described and analyzed using three
different modeling frameworks: timed automata (Uppaal), colored Petri nets and synchronous dataflow.
None of the models in \cite{Octopus08} incorporated uncertainty and in particular it was assumed that the arrival
times of new jobs are known in advance.
In reality, of course, the arrival time of new printer jobs is typically unknown (arrival times are the most significant
source of uncertainty in this application domain).
In this paper we use Uppaal Tiga to generate optimal scheduling strategies for scenarios in which the
arrival times of certain jobs is unknown.
Previous industrial applications of Uppaal Tiga deal with the synthesis of controllers \cite{JessenRLD07,CassezJLRR09}.
The present paper describes the first application to an industrial scheduling problem.

The rest of this paper is organized as follows.
Section~\ref{Case-study} recalls the printer case study and the Uppaal model from \cite{Octopus08}.
Section~\ref{Problem_description_and_Uppaal_Tiga} describes how uncertainty of arrival times can be modeled and analyzed
using Uppaal Tiga.
Section \ref{Results} presents an overview of the results that we obtained using Uppaal Tiga and discusses how these results
can be used to improve real printers/copiers. Finally, Section \ref{Conclusion}  gives some conclusions and directions for further work.
We assume that the reader has some knowledge of timed automata and Uppaal (see \cite{BehrmannDL04} for a tutorial).
The Uppaal Tiga models described in this paper are available on-line at the URL
\url{http://www.mbsd.cs.ru.nl/publications/papers/fvaan/TigaOce/}.

\section{Case Study}
\label{Case-study}
In this section, we introduce the Oc\'e case study and the Uppaal model for it.  The material in this section, which has been taken from \cite{Octopus08},
will be the starting point for the application of Uppaal Tiga, which we will describe in the subsequent sections of this paper.
The main challenge in the case study is to compute efficient schedules for printer/copier jobs that minimize latency and maximize throughput.

\subsection{Architecture}
\label{sec:Generic_Architecture} 
\begin{figure}[ht]
\begin{center}
\includegraphics[width = 12cm]{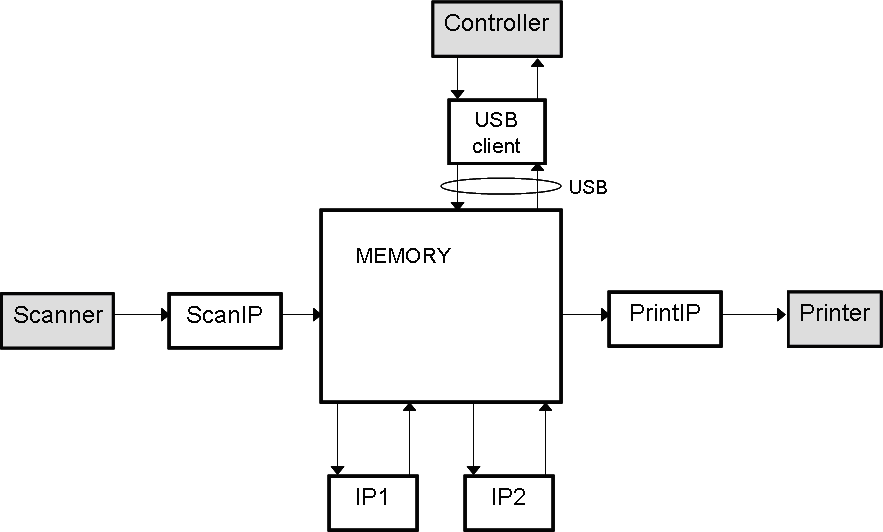}
\caption{Architecture of Oc\'e system}
\label{fig:Generic_architecture}
\end{center}
\end{figure}

Oc\'e systems perform a variety of image processing functions on digital documents in addition to scanning, copying and printing. Apart from local use for scanning and copying, users can also remotely use the system for image processing and printing. A generic architecture of the system studied in this paper is shown in Fig. \ref{fig:Generic_architecture}.
The system has two ports for input: Scanner and Controller. Users locally come to the system to submit jobs at the Scanner and remote jobs enter the system via the Controller. These jobs use the image processing (IP) components (ScanIP, IP1, IP2, PrintIP), and system resources such as memory and a USB bus for executing the jobs. Finally, there are two places where the jobs leave the system: Printer and Controller. 

The IP components can be used in different combinations depending on how a document is requested to be processed by the user. Hence this gives rise to different use cases of the system, that is, each job may use the system in a different way. The list of components used by a job defines the \emph{datapath} for that job. Some examples of datapaths are:
\\\\
- {\bf DirectCopy:} Scanner \( \leadsto \) ScanIP \( \leadsto \) IP1 \( \leadsto \) IP2  \( \leadsto \) USBClient, PrintIP\\
- {\bf ScanToStore:} Scanner \( \leadsto \) ScanIP \( \leadsto \) IP1 \( \leadsto \) USBClient\\
- {\bf ScanToEmail:} Scanner \( \leadsto \) ScanIP \( \leadsto \) IP1 \( \leadsto \) IP2 \( \leadsto \) USBClient\\
- {\bf ProcessFromStore:} USBClient \( \leadsto \) IP1 \( \leadsto \) IP2 \( \leadsto \) USBClient\\
- {\bf PrintWithProcessing:} USBClient \( \leadsto \) IP2 \( \leadsto \) PrintIP 
\\\\
Here $A \leadsto B$ means that the start of processing by $A$ should precede the start of processing by $B$.
In the {\it DirectCopy} datapath, a job is processed in order by the components Scanner, ScanIP, IP1, and IP2,
and then simultaneously sent to the Controller via the USBClient and to the printer through PrintIP.
The interpretation of the remaining datapaths is similar.

It is not mandatory that components in a datapath process a job sequentially: the design of the system allows for a certain degree of parallelism.
Scanner and ScanIP, for instance, may process a page in parallel.
This is because ScanIP processes the output of the scanner on a line-by-line basis (``streaming'') and has the same throughput as the Scanner.
However, due to the characteristics of the different components, some additional constraints are imposed.
Because of the nature of the image processing function that IP2 performs, IP2 can start processing a page only after IP1 has completed processing it.
The dependency between ScanIP and IP1 is different. IP1 processes the output of ScanIP in streaming mode and has a higher throughput than ScanIP.
Hence IP1 may start processing the page while ScanIP is processing it, with a certain delay due to the higher throughput of IP1.

In addition to the image processing components, two other system resources are memory and USB bandwidth. Execution of a job is only allowed if the entire memory required for completion of the job is available (and allocated) before its execution commences. Each component requires a certain amount of memory for its task and this can be released once computation has finished and no other component needs the information. Another critical resource is the USB. This bus has limited bandwidth and serves as a bridge between the USBClient and the memory. The bus may be used both for uploading and for downloading data. At most one job may upload data at any point in time, and similarly at most one job may download data. Uploading and downloading may take place concurrently. We assume that the transmission rate for both concurrent and non-concurrent bus usage is constant. 

\subsection{Model}
\label{sec:Explanation_of_the_Model}
In the Uppaal model, each use case and each resource is described by an automaton, except for memory which is simply modeled as a shared variable. 

\begin{figure}[ht]
\begin{center}
\includegraphics[width = 10 cm]{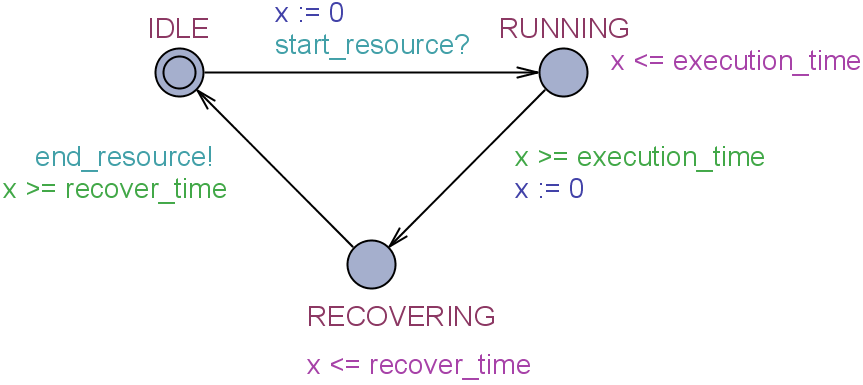}
\caption{Component template}
\label{fig:componentTemplate}
\end{center}
\end{figure}

All image processing components, as well as the USB, follow the same behavioral pattern, displayed in Fig.~\ref{fig:componentTemplate}.
Initially a component is in idle mode. As soon as the component is claimed by a job, it enters the running mode.
An integer variable {\sf execution\_time} specifies how long the automaton stays in this mode.
After this period has elapsed, the automaton jumps to the recovery mode, and stays there for {\sf recover\_time} time units.
The template of Fig.~\ref{fig:componentTemplate} is parameterized by channel names {\sf start\_resource} and {\sf end\_resource},
which mark the start and termination of a component, and integer variables {\sf execution\_time} and {\sf recover\_time},
which describe the timing behavior.

\begin{figure}[ht]
\begin{center}
\includegraphics[width = 14 cm]{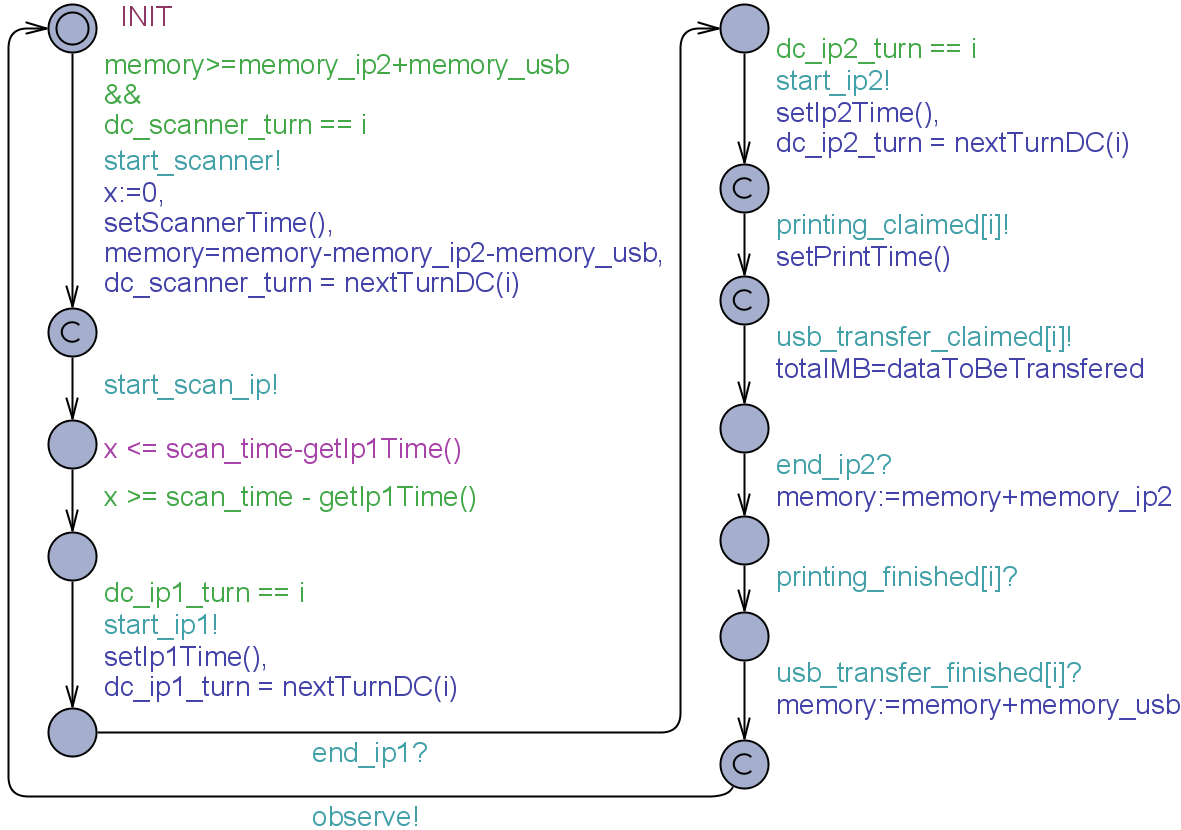}
\caption{\emph{DirectCopy} template}
\label{fig:directCopyTemplate}
\end{center}
\end{figure}

\begin{figure}[ht]
\begin{center}
\includegraphics[width = 10cm]{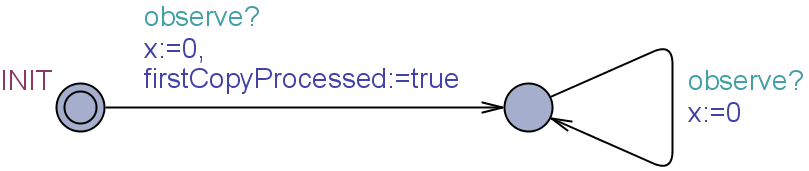}
\caption{Observer template}
\label{fig:observerTemplate}
\end{center}
\end{figure}

Each use case is modeled using a separate automaton.
As an example, the automaton for the {\it DirectCopy} use case is depicted in Fig.~\ref{fig:directCopyTemplate}.
A job may only claim the first component from its datapath when enough memory is available for all the processing in the datapath.
This figure shows the way memory allocation and release is modeled.
At the moment a component is claimed, the use case automaton specifies its execution time.
For reasons of space we have not included the definition of functions like {\sf setScannerTime()} in this paper; for the full definitions
we refer to the Uppaal model available at \url{http://www.mbsd.cs.ru.nl/publications/papers/fvaan/TigaOce/}.
The figure illustrates, also, the way we modeled the parallel activities of IP2, USBClient and PrintIP.
After finishing with the last component in the datapath, the automaton loops back to the {\sf INIT} location so that a new job can start.
This loop models a queue of {\it DirectCopy} jobs.

The {\it Observer} automaton, depicted in Fig.~\ref{fig:observerTemplate}, is used to measure the throughput of jobs.
Every time a {\it DirectCopy} job finishes, it synchronizes with an {\it Observer} instance, via the {\sf observe} channel,
to reset the observer clock {\sf x}. The {\it Observer} automaton has two locations because the first job being processed
can have a different throughput requirement than the subsequent jobs.
When the first job has been processed the flag {\sf firstCopyProcessed} is set to {\sf true}.
Section~\ref{Problem_description_and_Uppaal_Tiga} describes how winning conditions can be expressed using {\it Observer} instances. 


In order to reduce the state space, our model also incorporates a few of the scheduling rules that are commonly used in practice.
However, we only included rules that do not rule out optimal schedules.  A first scheduling rule that we included
is \emph{non-overtaking}, which states that different instances of the same use case may not overtake each other.
As described in \cite{BehrmannBHM05}, non-overtaking reduces the state space without losing optimal schedules.
Another rule that we included in our model is \emph{non-laziness} \cite{AAM06}.
Here the idea is that whenever a job needs a resource and this resource is available, the resource is allocated immediately, either to this
job or to another one.
In general, non-laziness may rule out optimal schedules, but we only introduce non-laziness for resources for which there is no competition.
In a setting where there is no competition for memory, this reduces the state space without losing optimal schedules.
Nonlaziness can be implemented in Uppaal by making the channels for claiming a resource urgent.


\section{Uppaal Tiga and Model Adaptation}
\label{Problem_description_and_Uppaal_Tiga}
In this section, we first describe a realistic scenario that illustrates the importance of unpredictable jobs, that is, jobs with uncertain arrival times. Then we investigate how Uppaal and Uppaal Tiga can be used to model these unpredictable jobs. Based on the results of this investigation we present an adaptation of the model described in Section~\ref{Case-study}. 

\subsection{A Realistic Scenario}
\label{sec:Realistic_Scenario}
The system described in Section~\ref{Case-study} has several different datapaths that use the system components in different ways.
In practice, a common scenario is that the printer/copier processes a series of {\it DirectCopy} jobs and at some time a
{\it PrintWithProcessing} job arrives.  One may think of a person making a direct copy of a document and someone else starting a
print job on the same machine from a remote location, at a time that is not known in advance.
The {\it PrintWithProcessing} job should be served in a reasonable time. However the {\it DirectCopy} jobs should not be delayed
for too long. Therefore we would like to find a strategy that can deal with the unpredictable nature of the {\it PrintWithProcessing}
job and still ensure an acceptable trade-off between the throughput of the {\it PrintWithProcessing} and the {\it DirectCopy} jobs.  

\subsection{Unpredictable Jobs in Uppaal}
\label{Unpredictable_Jobs_in_Uppaal}
Uncertainty of job arrival times can be modeled in Uppaal using non-determinism.
However, when we ask Uppaal to compute an optimal schedule, Uppaal will chose the arrival times of jobs in such a way that this allows
for the optimal schedule.  Since the Uppaal tool is based on timed automata, rather than timed game automata, we cannot introduce
a distinction between controllable and uncontrollable delays.
 
A simple scenario that demonstrates this problem is described here. Suppose we have two {\it PrintWithProcessing} jobs where one has a predictable arrival time and the other job is unpredictable in terms of arrival time. The trade-off property specifies that when the unpredictable job arrives it has a very low serve time compared to the predictable job. Both jobs need access to the {\it USBclient}, which is the first component in the data path of both jobs. This makes the {\it USBclient} the critical resource. If the unpredictable job needs a critical resource, Uppaal will ensure that this critical resource is not claimed by the predictable job in the optimal schedule. The reason for this is that the low serve time of the unpredictable job has to be satisfied quickly while the predictable job can be postpone without violating its serve time. Thus the strategy behind the optimal schedule has some knowledge about the \emph{future} arrival time of an unpredictable job in order to ensure that the critical resource is not claimed by a predictable job. 

\subsection{Unpredictable Jobs in Uppaal Tiga}
As described earlier, Uppaal Tiga defines a two player game with on the one side the controller (in charge of the controllable edges)
and on the other side the environment (in charge of the uncontrollable edges).
This strict separation between controller and environment can be used to generate strategies that do not depend on any knowledge of
the arrival time of unpredictable jobs. Intuitively we can see Uppaal as having one player, the controller, that has control over all edges.
In Uppaal Tiga we can use the concept of a game to separate control into two players.
The idea is that the player representing the environment has full control over the arrival time of unpredictable jobs.
The goal of the environment is to prevent the controller from satisfying the control objective by activating the unpredictable
jobs at the worst possible time. In the simple scenario described in the previous paragraph this would be at the time when the predictable
job has claimed the critical {\it USBclient} resource. The controller has to find a strategy to satisfy the trade-off constraint despite
the efforts of the environment. This prevents the controller strategy from having any knowledge about the arrival time of the unpredictable jobs.
Thus Uppaal Tiga can be used to investigate scenarios, like the one described in Section~\ref{sec:Realistic_Scenario}, that deal with unpredictable jobs.       
\subsection{Model Adaptation}
\label{sec:model-adaptation}

\begin{figure}[ht]
\begin{center}
\includegraphics[width = 16cm]{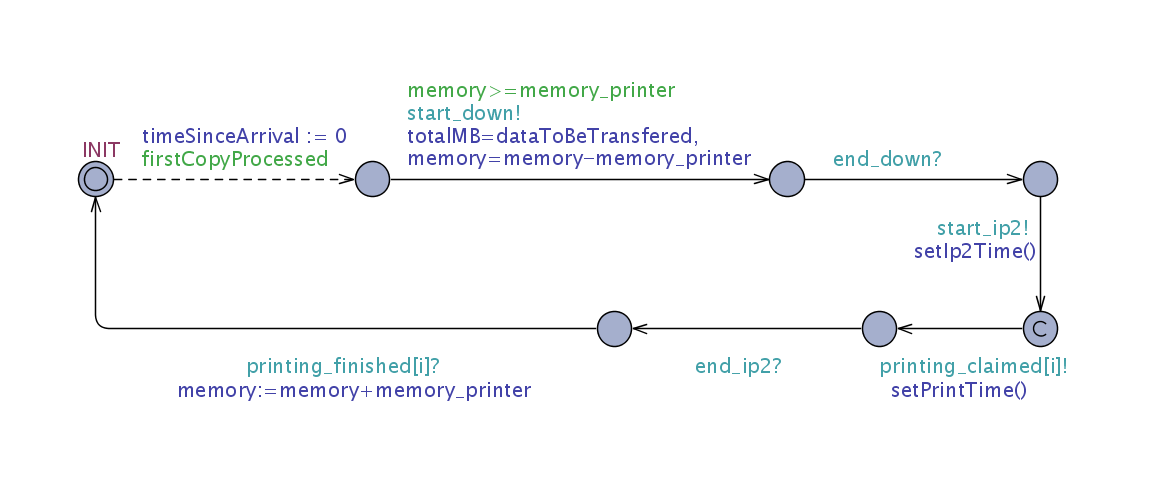}
\vspace{-1cm}
\caption{The automaton of {\it PrintWithProcessing} job in Tiga}
\label{fig:PrintWithProcessing}
\end{center}
\end{figure}

In order to reflect the unpredictable arrival time of a {\it PrintWithProcessing} job, the first transition of the corresponding automaton
is made uncontrollable, as shown in Figure \ref{fig:PrintWithProcessing}. This automaton represents a looping {\it PrintWithProcessing}
by having the last transition pointing back to the {\sf INIT} location. The clock {\sf TimeSinceArrival} measures the serve time of the
current {\it PrintWithProcessing} job and is reset upon arrival of a new job. According to the observer automaton discussed in
Section~\ref{sec:Explanation_of_the_Model}, the guard {\sf firstCopyProcessed} of the first transition ensures that the
{\it PrintWithProcessing} job arrives after the printer has processed the first {\it DirectCopy} job. 

With our adaptation of the model, it is easy to specify the winning condition of the strategy being generated. We use two looping instances of
{\it DirectCopy} jobs (DC) and one looping instance of a {\it PrintWithProcessing} job (DP). The actual number of processed jobs is unbounded,
but we make the assumption that there are only two DC and one DP job at the same time.
If there is only one concurrent DP job, two instances of DC jobs are enough to keep the scanner busy all the time.
The winning condition can be expressed as follows:
\begin{verbatim}
control:A[] (DC_OBSERVER.INIT imply DC_OBSERVER.x <= FIRST_DC_TIME)  &&
	(!DC_OBSERVER.INIT imply DC_OBSERVER.x <= DC_TIME)  &&
	(!DP0.INIT imply DP0.timeSinceArrival <= DP_TIME)
\end{verbatim}
As described in Section~\ref{sec:Explanation_of_the_Model}, the first line states that the first DC job has to be served within
{\tt FIRST\_DC\_TIME} time units.
We can chose an arbitrary value here as long as the first job can be processed within that time, without changing the result.
The only thing that is expressed here is that the first job eventually finishes.
After the first DC job has been served, the DP jobs may arrive at unpredictable times.
The variables {\tt DC\_TIME} and {\tt DP\_TIME} together represent a trade-off between the throughput of DC and the serve time of DP jobs.
The second and third line express that such a trade-off should be satisfied. 

We tried different combinations of values of the serve time of DCs ({\tt DC\_TIME}) and the serve time of DP ({\tt DP\_TIME})
to find out the trade-off where a strategy exists.
The idea is, for a fixed value of the serve time of one kind of job, to determine the minimal value of the serve time for the other kind of job.

\section{Results} \label{Results}
In this section, we report on the results obtained with Uppaal Tiga for the Oc\'{e} case study.
	
\subsection{Optimal Strategies}
\begin{figure}[ht]
				\begin{center}
					\includegraphics[width = 15cm]{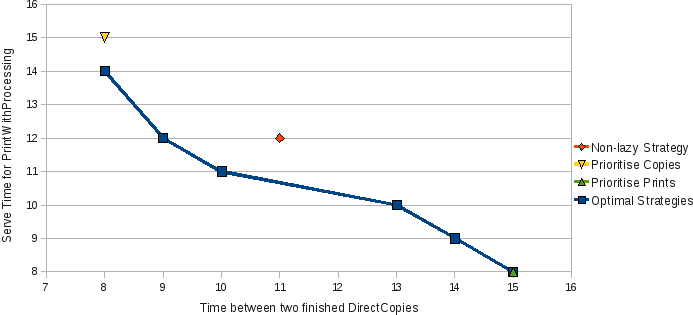}
					\caption{Pareto Frontier Representing Optimal Strategies \& Timings for Fixed Strategies}
					\label{fig:result}
				\end{center}
			\end{figure}

We found six optimal strategies, each providing a different trade-off between the throughput of \emph{DirectCopies}
and the maximal serve time of \emph{PrintWithProcessing} jobs.
These optimal strategies are represented as a \emph{Pareto frontier} in Figure~\ref{fig:result}.
Note that, since constants {\tt DC\_TIME} and {\tt DP\_TIME} may only take integer values, no formal meaning
can be attached to points on the frontier except those with integer coordinates.
The minimal serve time for \emph{PrintWithProcessing} jobs is $7$ in the case where no other jobs are present. However,
there is no strategy achieving this under the condition that the \emph{PrintWithProcessing} jobs should finish eventually.
For \emph{PrintWithProcessing} a strategy for achieving the lowest possible serve time can be found. This strategy leads to
a throughput for \emph{DirectCopies} with is more than twice as low as the optimal one.
			
There are some gaps in the data-points. For example, for a maximal time between two finished \emph{DirectCopy}
jobs of $11$ and $12$ there is no corresponding time on the y-axis. This is because the values for both are $11$,
which is the same as with a lower maximal time between two finished \emph{DirectCopy} jobs of $10$.
So these strategies are always worse and therefore we did not depict them in the diagram.

Uppaal Tiga also offers the possibility to export the generated strategies as a list of rules.
However, the strategies created by Tiga are not suitable to be built directly into a real printer controller.
The first reason for this is that the set of rules is huge and contains thousands of rules.
The second reason is that the rules contain information about the state of all automata,
including for instance the auxiliary agent automata which are not present in a real system.
Also, the strategy only works for a situation with a fixed number of jobs and is therefore not generally applicable.

\subsection{Simple Strategies}
Although the strategies generated by Uppaal Tiga cannot directly be used within a real printer controller,
they can still be helpful for improving the scheduling of a printer.
In practice, the scheduling strategy of a printer should be very simple.
What we can do is to implement simple strategies in our model and compare the timings for them with the optimal ones found by Uppaal Tiga.
In Figure~\ref{fig:result} three data-points representing the performance of fixed strategies are given.
		
The first very simple strategy is to make all resources non-lazy (that is, all channels urgent).
In theory this does not entirely fix the strategy because if two jobs are waiting for the same resource there is a nondeterministic
choice which jobs gets the resource at the moment it becomes available. However for the small number of jobs in our scenario this situation is not relevant and we get a fixed strategy. As one can see, the timings for this simple strategy are worse than the optimal ones.
		
We also considered two strategies that prioritize one kind of job.
For prioritising \emph{DirectCopies} we only allow a \emph{PrintWithProcessing} job to use IP2 when the scanner is not in use.
With this strategy the throughput for the \emph{DirectCopies} is optimal but the serve time of the \emph{PrintWithProcessing} jobs is
higher than with the optimal strategy. This is because the scanner uses more time than IP2 and consequently there are situations where
using IP2 is not a problem because it will be free again at the moment the scanner is finished.
	
In the strategy that gives priority to \emph{PrintWithProcessing} jobs, a \emph{DirectCopy} may not use IP2 if a \emph{PrintWithProcessing} job
arrives at the USB bus. But this is not sufficient, since there is more than one instance of \emph{DirectCopy}.
It may happen that a \emph{DirectCopy} is still using the printer when the \emph{PrintWithProcessing} job finishes using IP2
and wants to use the printer.
To solve this issue, we added another constraint that a \emph{DirectCopy} may only use IP2 if another instance of \emph{DirectCopy}
is not using the printer. While this decreases the average throughput, for the situation where we want to prioritise another kind of job, never two \emph{DirectCopies} are printed without another kind of job between them. So we get a strategy which is as good as the optimal one for this case.

\subsection{Results for an Extended Model}
		
We have also experimented with an extended version of the model presented in Section~\ref{Case-study}. This extended model includes many features of an Oc\'{e} printer which is currently in the design phase\footnote{Because of space limit we cannot add a full description of the model, but we invite the reader to have a look at the model available at \url{http://www.mbsd.cs.ru.nl/publications/papers/fvaan/TigaOce/}.}. One example of such a feature is the memory bus, which is the interface between the components and the memory. Memory management was also refined, for instance, print jobs access different memory locations than scan jobs. Besides these, the numerical values were similar to those used in reality.
		
The most important change w.r.t.\ the model of Section~\ref{Case-study} is the modeling of the scheduling rules employed by the printer's controller. One example of such a rule is a memory bus arbitration, which is based on a priority mechanism between the components. Another example is the rule used for releasing the print job memory when the Print and the USB Upload actions are delayed. With these rules, the model essentially becomes fully deterministic. However, in Uppaal (Tiga) no partial order reduction mechanism is implemented. As a result of large numbers of (confluent) synchronization transitions, we could not analyze scenarios with a large number of jobs and big amount of memory available. We mention here one example of a scenario which we could analyze. We have used two different datapaths: ScanToEmail and PrintWithProcessing where Print is added in parallel with USBClient. In this scenario two looping ScanToEmail jobs are present in the system together with one looping PrintWithProcessing job. Non-lazy scheduling is used, memory can buffer one print job and one scan job and 10 Scan to Email jobs are processed. For this scenario, Uppaal Tiga could compute the fastest schedule (strategy) when one job type is prioritized.  The only simplification we made in this model is that we reduced the memory size so that the system can store fewer jobs.
If we remove some of the scheduling rules so that the model becomes (essentially) nondeterministic again, Uppaal Tiga is no longer able to handle the model.

\section{Conclusions \& Future Work}
\label{Conclusion}
In this paper, we have applied Uppaal Tiga to automatically compute adaptive scheduling strategies
for an industrial case study dealing with a state-of-the-art image processing pipeline of a printer.
As far as we know, this is the first application of timed automata technology to an industrial
scheduling problem with uncertainty in job arrivals.

Despite our promising initial results, the problem to automatically synthesize \emph{practical} scheduling
strategies for this application domain is still wide open.
In some other applications, the strategies synthesized by Tiga have been used directly in the
generation of control software \cite{JessenRLD07,CassezJLRR09}.
This is currently not possible for the printer case study described in this paper:
the strategies produced by Tiga (albeit memoryless) are really big and contains
tens of thousands of rules.
Even though there appears to be room to improve Tiga so that it generates
smaller strategies, direct use of Tiga strategies for scheduling of printers
will probably not be practical for the years to come.
We had to restrict our analysis to a scenario with a fixed continuous stream of
direct copy jobs and a single uncontrollable direct print job, simply because
Tiga cannot handle more uncontrollable jobs.
Also, the model that we described in Section~\ref{Case-study} is a simplification of the
realistic and more complex models that we constructed in the context of the Octopus project.
For realistic models, the generated strategies will even be much bigger
and far beyond the tight constraints on CPU and memory usage of todays printer controllers.

Nevertheless, we believe that we are close to the point at which Uppaal Tiga can already be useful in the actual
design of data path controllers.  These controllers typically consists of a relatively small number of simple
rules that determine which resource is allocated to which job
(for instance: job and resource priorities, FCFS for jobs with same priority, greedy resource allocation,..).
Due to state space explosion, Tiga will not be able to handle unconstrained realistic printer models.
However, our experiments show that Tiga is able to deal with realistic printer models of which the
state space is constrained by a few of the simple scheduling rules that Oc\'{e} wants to use anyway.
By applying Tiga to (downsized) versions of printer models and replaying the strategies in the simulator,
we may be able to come up with simple new control rules that are implementable on real printers.
Tiga may also give an indication of how close the implemented rules are from the optimum.

\bibliographystyle{eptcs}
\bibliography{abbreviations,dbase}

\newcommand{\SortNoop}[1]{}
\begin{thebibliography}{10}
\providecommand{\bibitemstart}[1]{\bibitem{#1}}
\providecommand{\bibitemend}{}
\providecommand{\bibliographystart}{}
\providecommand{\bibliographyend}{}
\providecommand{\url}[1]{\texttt{#1}}
\providecommand{\urlprefix}{Available at }
\providecommand{\bibinfo}[2]{#2}
\bibliographystart

\bibitemstart{AAM06}
\bibinfo{author}{Y.~Abdedda\"{\i}m}, \bibinfo{author}{E.~Asarin} \&
  \bibinfo{author}{O.~Maler} (\bibinfo{year}{2006}):
  \emph{\bibinfo{title}{Scheduling with timed automata}}.
\newblock {\sl \bibinfo{journal}{Theor. Comput. Sci.}}
  \bibinfo{volume}{354}(\bibinfo{number}{2}), pp. \bibinfo{pages}{272--300}.
\bibitemend

\bibitemstart{AD94}
\bibinfo{author}{R.~Alur} \& \bibinfo{author}{D.L. Dill}
  (\bibinfo{year}{1994}): \emph{\bibinfo{title}{A theory of timed automata}}.
\newblock {\sl \bibinfo{journal}{Theoretical Computer Science}}
  \bibinfo{volume}{126}, pp. \bibinfo{pages}{183--235}.
\bibitemend

\bibitemstart{AMETISTfinal}
\bibinfo{author}{AMETIST} (\bibinfo{year}{2007}).
\newblock \emph{\bibinfo{title}{Final Project Report}}.
\newblock \urlprefix\url{http://www.cs.ru.nl/F.Vaandrager/AMETIST/final.pdf}.
\newblock \bibinfo{note}{Deliverable from the European project IST-2001-35304
  Advanced Methods for Timed Systems (AMETIST)}.
\bibitemend

\bibitemstart{BehrmannBHM05}
\bibinfo{author}{G.~Behrmann}, \bibinfo{author}{E.~Brinksma},
  \bibinfo{author}{M.~Hendriks} \& \bibinfo{author}{A.~Mader}
  (\bibinfo{year}{2005}): \emph{\bibinfo{title}{Production Scheduling by
  Reachability Analysis - A Case Study}}.
\newblock In: {\sl \bibinfo{booktitle}{19th International Parallel and
  Distributed Processing Symposium (IPDPS 2005), CD-ROM / Abstracts
  Proceedings, 4-8 April 2005, Denver, CA, USA}}. \bibinfo{publisher}{IEEE
  Computer Society}.
\newblock
  \urlprefix\url{http://doi.ieeecomputersociety.org/10.1109/IPDPS.2005.363}.
\bibitemend

\bibitemstart{tiga_manual}
\bibinfo{author}{G.~Behrmann}, \bibinfo{author}{A.~Cougnard},
  \bibinfo{author}{A.~David}, \bibinfo{author}{E.~Fleury},
  \bibinfo{author}{K.G. Larsen} \& \bibinfo{author}{D.~Lime}
  (\bibinfo{year}{2007}).
\newblock \emph{\bibinfo{title}{{U}ppaal {T}iga User-manual}}.
\newblock \urlprefix\url{http://www.cs.aau.dk/~adavid/tiga/manual.pdf}.
\bibitemend

\bibitemstart{BehrmannDL04}
\bibinfo{author}{G.~Behrmann}, \bibinfo{author}{A.~David} \&
  \bibinfo{author}{K.G. Larsen} (\bibinfo{year}{2004}): \emph{\bibinfo{title}{A
  Tutorial on {U}ppaal}}.
\newblock In: \bibinfo{editor}{M.~Bernardo} \& \bibinfo{editor}{F.~Corradini},
  editors: {\sl \bibinfo{booktitle}{Formal Methods for the Design of Real-Time
  Systems, International School on Formal Methods for the Design of Computer,
  Communication and Software Systems, SFM-RT 2004, Bertinoro, Italy, September
  13-18, 2004, Revised Lectures}}, {\sl \bibinfo{series}{Lecture Notes in
  Computer Science}} \bibinfo{volume}{3185}. \bibinfo{publisher}{Springer}, pp.
  \bibinfo{pages}{200--236}.
\bibitemend

\bibitemstart{CDFLL05}
\bibinfo{author}{F.~Cassez}, \bibinfo{author}{A.~David},
  \bibinfo{author}{E.~Fleury}, \bibinfo{author}{K.G. Larsen} \&
  \bibinfo{author}{D.~Lime} (\bibinfo{year}{2005}):
  \emph{\bibinfo{title}{Efficient On-the-Fly Algorithms for the Analysis of
  Timed Games}}.
\newblock In: \bibinfo{editor}{Mart\'{\i}n Abadi} \& \bibinfo{editor}{Luca
  de~Alfaro}, editors: {\sl \bibinfo{booktitle}{CONCUR 2005 - Concurrency
  Theory, 16th International Conference, CONCUR 2005, San Francisco, CA, USA,
  August 23-26, 2005, Proceedings}}, {\sl \bibinfo{series}{Lecture Notes in
  Computer Science}} \bibinfo{volume}{3653}. \bibinfo{publisher}{Springer}, pp.
  \bibinfo{pages}{66--80}.
\newblock \urlprefix\url{http://dx.doi.org/10.1007/11539452_9}.
\bibitemend

\bibitemstart{CassezJLRR09}
\bibinfo{author}{F.~Cassez}, \bibinfo{author}{J.J. Jessen},
  \bibinfo{author}{K.~G. Larsen}, \bibinfo{author}{J.-F. Raskin} \&
  \bibinfo{author}{P.-A. Reynier} (\bibinfo{year}{2009}):
  \emph{\bibinfo{title}{Automatic Synthesis of Robust and Optimal Controllers -
  An Industrial Case Study}}.
\newblock In: \bibinfo{editor}{Rupak Majumdar} \& \bibinfo{editor}{Paulo
  Tabuada}, editors: {\sl \bibinfo{booktitle}{Hybrid Systems: Computation and
  Control, 12th International Conference, HSCC 2009, San Francisco, CA, USA,
  April 13-15, 2009. Proceedings}}, {\sl \bibinfo{series}{Lecture Notes in
  Computer Science}} \bibinfo{volume}{5469}. \bibinfo{publisher}{Springer}, pp.
  \bibinfo{pages}{90--104}.
\newblock \urlprefix\url{http://dx.doi.org/10.1007/978-3-642-00602-9_7}.
\bibitemend

\bibitemstart{DB00}
\bibinfo{author}{A.J. Davenport} \& \bibinfo{author}{J.C. Beck}
  (\bibinfo{year}{2000}).
\newblock \emph{\bibinfo{title}{A Survey of Techniques for Scheduling with
  Uncertainty}}.
\newblock \bibinfo{note}{Unpublished manuscript}.
\bibitemend

\bibitemstart{Octopus08}
\bibinfo{author}{G.~Igna}, \bibinfo{author}{V.~Kannan},
  \bibinfo{author}{Y.~Yang}, \bibinfo{author}{T.~Basten},
  \bibinfo{author}{M.~Geilen}, \bibinfo{author}{F.~Vaandrager},
  \bibinfo{author}{M.~Voorhoeve}, \bibinfo{author}{S.~de Smet} \&
  \bibinfo{author}{L.~Somers} (\bibinfo{year}{2008}):
  \emph{\bibinfo{title}{Formal Modeling and Scheduling of Datapaths of Digital
  Document Printers}}.
\newblock In: {\sl \bibinfo{booktitle}{Proceedings Sixth International
  Conference on Formal Modeling and Analysis of Timed Systems (FORMATS 2008),
  {\rm September 15-17, 2008, Saint-Malo, France}}}, {\sl
  \bibinfo{series}{Lecture Notes in Computer Science}} \bibinfo{volume}{5215}.
  \bibinfo{publisher}{Springer Berlin / Heidelberg}, pp.
  \bibinfo{pages}{169--186}.
\newblock
  \urlprefix\url{http://www.cs.ru.nl/ita/publications/papers/fvaan/Octopus08.h%
tml}.
\bibitemend

\bibitemstart{JessenRLD07}
\bibinfo{author}{J.J. Jessen}, \bibinfo{author}{J.~Rasmussen},
  \bibinfo{author}{K.G. Larsen} \& \bibinfo{author}{A.~David}
  (\bibinfo{year}{2007}): \emph{\bibinfo{title}{Guided Controller Synthesis for
  Climate Controller Using Uppaal Tiga}}.
\newblock In: \bibinfo{editor}{Jean-Fran\c{c}ois Raskin} \&
  \bibinfo{editor}{P.~S. Thiagarajan}, editors: {\sl \bibinfo{booktitle}{Formal
  Modeling and Analysis of Timed Systems, 5th International Conference, FORMATS
  2007, Salzburg, Austria, October 3-5, 2007, Proceedings}}, {\sl
  \bibinfo{series}{Lecture Notes in Computer Science}} \bibinfo{volume}{4763}.
  \bibinfo{publisher}{Springer}, pp. \bibinfo{pages}{227--240}.
\bibitemend

\bibitemstart{MPS95}
\bibinfo{author}{O.~Maler}, \bibinfo{author}{A.~Pnueli} \&
  \bibinfo{author}{J.~Sifakis} (\bibinfo{year}{1995}): \emph{\bibinfo{title}{On
  the Synthesis of Discrete Controllers for Timed Systems}}.
\newblock In: \bibinfo{editor}{E.W. Mayr} \& \bibinfo{editor}{C.~Puech},
  editors: {\sl \bibinfo{booktitle}{Proceedings STACS'95}}, {\sl
  \bibinfo{series}{Lecture Notes in Computer Science}} \bibinfo{volume}{900}.
  \bibinfo{publisher}{Springer-Verlag}, pp. \bibinfo{pages}{229--242}.
\bibitemend

\bibitemstart{MBS89}
\bibinfo{author}{K.N. McKay}, \bibinfo{author}{J.A. Buzacott} \&
  \bibinfo{author}{F.R. Safayeni} (\bibinfo{year}{1989}):
  \emph{\bibinfo{title}{The Scheduler's Knowledge of Uncertainty: The Missing
  Link}}.
\newblock In: \bibinfo{editor}{J.~Browne}, editor: {\sl
  \bibinfo{booktitle}{Knowledge Based Production Management Systems}}.
  \bibinfo{publisher}{Elsevier Science, New York}, pp.
  \bibinfo{pages}{171--189}.
\bibitemend

\bibitemstart{MSB98}
\bibinfo{author}{K.N. McKay}, \bibinfo{author}{F.R. Safayeni} \&
  \bibinfo{author}{J.A. Buzacott} (\bibinfo{year}{1998}):
  \emph{\bibinfo{title}{Job-shop scheduling theory: What is relevant?}}
\newblock {\sl \bibinfo{journal}{Interfaces}}
  \bibinfo{volume}{18}(\bibinfo{number}{4}), pp. \bibinfo{pages}{84--90}.
\bibitemend

\bibitemstart{MW99}
\bibinfo{author}{K.N. McKay} \& \bibinfo{author}{V.C.S. Wiers}
  (\bibinfo{year}{1999}): \emph{\bibinfo{title}{Unifying the Theory and
  Practice of Production Scheduling}}.
\newblock {\sl \bibinfo{journal}{Journal of Manufacturing Systems}}
  \bibinfo{volume}{18}(\bibinfo{number}{4}), pp. \bibinfo{pages}{241--255}.
\bibitemend

\bibliographyend
\end{thebibliography}

\end{document}